\documentclass[conference]{IEEEtran}
\IEEEoverridecommandlockouts
\usepackage{cite}
\usepackage{amsmath,amssymb,amsfonts}
\usepackage{algorithmic}
\usepackage{graphicx}
\usepackage{textcomp}
\usepackage{xcolor}
\usepackage{array}

\usepackage[normalem]{ulem}
\usepackage{listings}
\usepackage{soul}
\usepackage{balance}
\usepackage[labelfont=bf,textfont=bf,font=small]{subcaption} 
\usepackage[labelfont=bf,textfont=bf,font=small]{caption} 
\usepackage{pifont} 
\usepackage{multirow}
\usepackage{algorithm}

\usepackage{url}

\newcommand{\pname}[1]{{CODAG}{#1}}

\def\BibTeX{{\rm B\kern-.05em{\sc i\kern-.025em b}\kern-.08em
    T\kern-.1667em\lower.7ex\hbox{E}\kern-.125emX}}
\begin{document}
\title{CODAG: Characterizing and Optimizing Decompression Algorithms for GPUs}

\author{\IEEEauthorblockN{ Jeongmin Park}
\IEEEauthorblockA{\textit{UIUC} \\
jpark346@illinois.edu}

\and
\IEEEauthorblockN{Zaid Qureshi}
\IEEEauthorblockA{
  \textit{UIUC}\\
zaidq2@illinois.edu
}

\and
\IEEEauthorblockN{Vikram Sharma Mailthody}
\IEEEauthorblockA{%
  \textit{UIUC} \\
    vsm2@illinois.edu
}
\and
\IEEEauthorblockN{Andrew Gacek}
\IEEEauthorblockA{%
  \textit{UIUC} \\
  andrewg3@illinois.edu
}

\and
\IEEEauthorblockN{Shunfan Shao}
\IEEEauthorblockA{%
  \textit{UIUC} \\
shunfan2@illinois.edu
}

\and
\IEEEauthorblockN{Mohammad AlMasri}
\IEEEauthorblockA{%
  \textit{UIUC} \\
  almasri3@illinois.edu
}
\and

\IEEEauthorblockN{Isaac Gelado}
\IEEEauthorblockA{%
  \textit{NVIDIA} \\
  igelado@nvidia.com
}

\and
\IEEEauthorblockN{Jinjun Xiong}
\IEEEauthorblockA{%
  \textit{University at Buffalo} \\
 jinjun@buffalo.edu
}

\and
\IEEEauthorblockN{Chris J Newburn}
\IEEEauthorblockA{%
  \textit{NVIDIA} \\
  cnewburn@nvidia.com
}
\and

\IEEEauthorblockN{I-hsin Chung}
\IEEEauthorblockA{%
  \textit{IBM Research} \\
ihchung@us.ibm.com
}
\and

\IEEEauthorblockN{Michael Garland}
\IEEEauthorblockA{%
  \textit{NVIDIA}\\
mgarland@nvidia.com
}
\and

\IEEEauthorblockN{Nikolay Sakharnykh}
\IEEEauthorblockA{%
  \textit{NVIDIA} \\
  nsakharnykh@nvidia.com
}

\and
\IEEEauthorblockN{Wen-mei Hwu}
\IEEEauthorblockA{%
  \textit{UIUC / NVIDIA} \\
    w-hwu@illinois.edu
}

}

\maketitle
\begin{abstract}
Data compression and decompression have become vital components of big-data applications to manage the exponential growth in 
 the amount of data collected and stored.
Furthermore, big-data applications have increasingly adopted GPUs due to their high compute throughput and memory bandwidth.
Prior works presume that decompression is memory-bound and have dedicated most of the GPU's threads to data movement and adopted complex software techniques to hide memory latency for reading compressed data and writing uncompressed data. 
This paper shows that these techniques lead to poor GPU resource utilization as most threads 
end up waiting for the few decoding threads, exposing compute and synchronization latencies.

Based on this observation, we propose \pname{}, a novel and simple kernel architecture for high throughput decompression on GPUs.
~\pname{} eliminates the use of 
specialized groups of threads, 
frees up compute resources to increase the number of parallel decompression streams, and leverages the ample compute activities and the GPU's hardware scheduler to 
tolerate synchronization, compute, and memory latencies. Furthermore, \pname{} provides a framework for users to easily incorporate new decompression algorithms without being burdened with implementing complex optimizations to hide memory latency.
We validate our proposed architecture with three different
encoding techniques, RLE v1, RLE v2, and Deflate, 
and 
a wide range of large datasets from different domains. 
We show that \pname{} provides 13.46$\times$, 5.69$\times$, and 1.18$\times$ speed up for RLE v1, RLE v2, and Deflate, respectively, when compared to the state-of-the-art decompressors from NVIDIA RAPIDS.

\end{abstract}



\section{Introduction}
\label{sec:intro}

Compression has become a vital component of the data storage infrastructure due to unprecedented data size growth in 
big-data, high-performance computing (HPC), and data analytics applications~\cite{rapids,orc,pandas,nvcomp,nvcompblog,gdscompblog}.  
Compression aims to reduce the number of bits used to represent the data through various encoding techniques specialized for different data types and patterns. 
Compressed data must be first decompressed before it can be processed by compute operations.

As GPUs are increasingly 
used to meet the compute throughput demands of 
emerging big-data  workloads~\cite{application1, application2, application3}, data analytics pipelines typically read compressed data from the storage devices into the GPU memory 
and deploy one or more GPU kernels 
for decompression before performing query computation on the data. 
Prior works~\cite{hippogriffdb,gdscompblog} have shown this approach's benefits over storing and moving uncompressed data.
Thus, high-throughput decompression on the GPU is becoming a key requirement to avoid 
performance bottlenecks in these pipelines. 
Our profiling of data-dependent queries with the state-of-the-art GPU-enabled RAPIDS pipeline~\cite{rapids}, e.g., ``What is the average fare per trip for ride starting from Williamsburg?'' on the NYC taxicab dataset~\cite{nyctaxi}, a popular dataset for evaluating data analytics pipelines,  
shows that approximately 91\% of the total GPU execution time is consumed by the decompression kernel.

Prior works~\cite{all, gompress, tile_compression} have proposed various data-layout transformations on compressed datasets to reduce the GPU execution time spent on decompression. 
However, applying data-layout transformations 
on existing massive datasets
of modern big-data applications~\cite{bam, orc, hippogriffdb, parquet, criteo} is an expensive proposition, making decompression techniques that require such transformations impractical. 
Therefore, it's much more desirable for a decompression scheme to support the standard compressed data formats, e.g., ORC~\cite{orc}, and
not require data-layout transformations, a design goal that we aim to achieve in this work.


To better understand why decompression 
accounts for so much of the GPU execution time, 
we performed a detailed profiling study of the state-of-the-art GPU
decompressor in RAPIDS. 
As modern compression schemes support indexed access to multiple  chunks of compressed data that correspond to fixed-size chunks of uncompressed input data, 
the state-of-the-art decompressor  
exploits chunk-level parallelism by assigning compressed chunks to thread blocks. 
However, data dependencies still serialize the decoding operations within each chunk. Thus, the state-of-the-art decompressor 
dedicates a single thread, the leader thread, in each thread block to decode the compressed data and broadcast the decoded information to the remaining threads in the thread block. All threads then collaborate to execute the memory writing or copying operations based on the decoded information. The state-of-the-art decompressor further enhances the memory bandwidth utilization and hides more of the memory latency by dedicating one warp in each thread block to prefetch compressed data. 

By profiling the RAPIDS decompressor, 
we observe that (1) the memory-bandwidth-driven resource provisioning strategy in the state-of-the-art decompressor in fact fails to effectively hide compute operation, memory, and synchronization latencies; 
and (2) the root cause of the exposed latencies is the insufficient number of decoding (leader) threads that cannot take full advantage of the GPU's parallelism and provide the GPU warp schedulers with sufficient numbers of concurrently available instructions to hide
latencies.

Based on these key observations, we propose \pname{}, a modular and flexible framework for high-throughput decompression on GPUs. In~\pname{}, we assign compressed chunks to warps rather than thread blocks to drastically increase the exploitation of the GPU's parallelism to process more chunks at any given time, while maintaining the ability to exploit fine-grained parallelism within each chunk.
We refer to the assignment of each chunk to a warp as \emph{warp-level decompression}.
We believe that the warp-level decompression offers a number of benefits in \pname{}. First, warp-level decompression results in a much higher number of active decoding threads, 
which helps the hardware scheduler to effectively hide memory, synchronization, and decoding compute operation latencies. Second, since warp-level synchronization primitives have much lower latency and higher throughput than block-level synchronization, warp-level decompression reduces synchronization overhead.

On top of that, we further propose (1) a novel strategy for \pname{} to ensure that all on-demand reads or writes in each warp are performed on a cache-line granularity and are thus coalesced, making efficient use of the available GPU memory bandwidth;
(2) a clean abstraction for common operations during decompression with
the associated Application Programming Interface (API) functions as implemented in decompression kernels. 
This new abstraction provides the flexibility for programmers to easily incorporate their own encoding techniques into the decoding section of the \pname{} kernels while leveraging the rest of the kernel to achieve the desired level of latency tolerance and memory bandwidth efficiency.



We make the following key contributions in this paper.

\begin{itemize}
\item We conduct a systematic characterization study on the state-of-the-art GPU decompressor, identify the performance bottlenecks, and determine that decompression on modern GPUs is a compute-bound process.


\item We present \pname{}, a novel and superior GPU resource provisioning strategy for decompressors with various optimization techniques and abstraction. The proposed APIs can support common decompression operations so that users can easily integrate their compression algorithms into \pname{} with low programming effort.

\item We demonstrate \pname{}'s effectiveness and flexibility by implementing three encoding techniques,
i.e., RLE v1, RLE v2, and Deflate. Results based on the NVIDIA A100 GPUs show that \pname{} achieves 13.46$\times$, 5.69$\times$, and 1.18$\times$ speedups over a state-of-the-art decompressor, respectively.


\end{itemize}


\section{Background}
\label{sec:background}

In this section, we provide an overview of common compression encoding techniques and how their decompression algorithms have been traditionally parallelized.
We then briefly explain why the {resource provisioning strategies used in the state-of-the-art GPU decompressors limit the performance and} efficiency of
decompression on GPUs.

\subsection{Traditional Encoding Techniques}
\label{sec:comp_techniques}
Generally, compressors
take a contiguous sequence of bytes and attempt to encode that sequence in as few symbols as possible.
The primary measure of merit of a compressor is the ratio of the size of the uncompressed data over that of the compressed data, referred to as the compression ratio.
To maximize compression ratio, compressors use an arsenal of encoding techniques each of which targets different data types and patterns. 
Here, we describe some of the most common encoding techniques. 

The run-length encoding (RLE) technique takes advantage of
the same byte or value being repeated in a contiguous sequence, called a~\emph{run}. The RLE v1 encoding scheme encodes each \emph{run} 
with two values: a literal and the length of the run. 

Delta encoding provides a high compression ratio when there is a contiguous sequence of values and the differences (deltas) between adjacent values are small. 
Delta encoding stores the starting value {and the number of values in the sequence} followed by the deltas to create a compressed data representation. 
The RLE v2 encoding scheme leverages delta encoding in addition to RLE to capture more data patterns and further improve the compression ratio over RLE v1.

Dictionary-based encoding generates 
compact encoding 
when sequences of values are repeated between different sections of the uncompressed data. Given a sequence of bytes, the decompressor can use a dictionary to look up the sequence to see if it has been encountered in the past. 
If so, the repeated instance can be encoded with 2 values, an offset into the dictionary, and the length of the repeated sequence.

In dictionary encoding, some sequences are repeated more than others. Thus, some references (offset/length pairs) appear more frequently than others \cite{deflate} in the compressed stream. The Deflate encoding scheme sorts the literals and the length-offset pairs according to their frequency into a Huffman tree \cite{huffman} so that the number of bits used for each is inversely proportional to their frequency, thus minimizing the total number of bits used.



\begin{figure*}[t]
\centering
{
    \begin{subfigure}[b]{0.48\textwidth}
         \centering
         \includegraphics[width=0.9\textwidth]{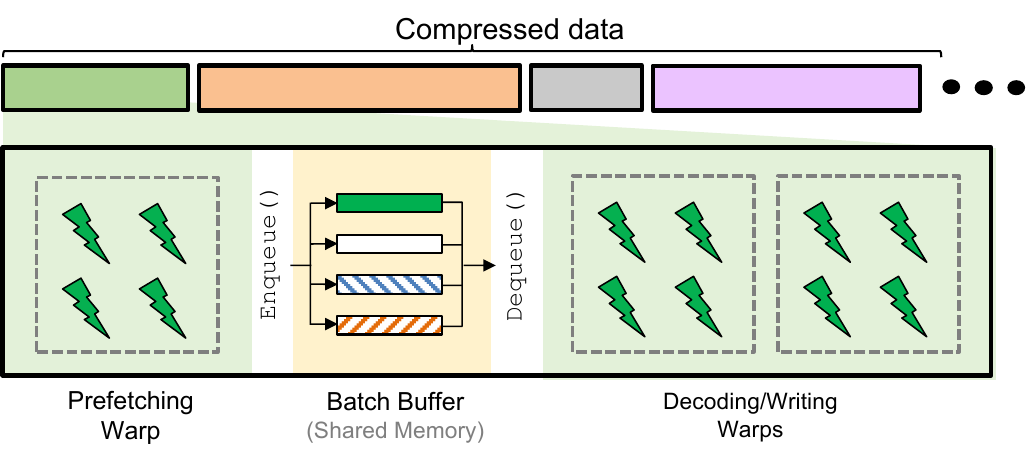}
         \vspace{-1ex}
        \caption{State-of-the-art} 
        \label{fig:baselinedesign}
     \end{subfigure}
     \hfill
     \begin{subfigure}[b]{0.48\textwidth}
         \centering
         \includegraphics[width=0.9\textwidth]{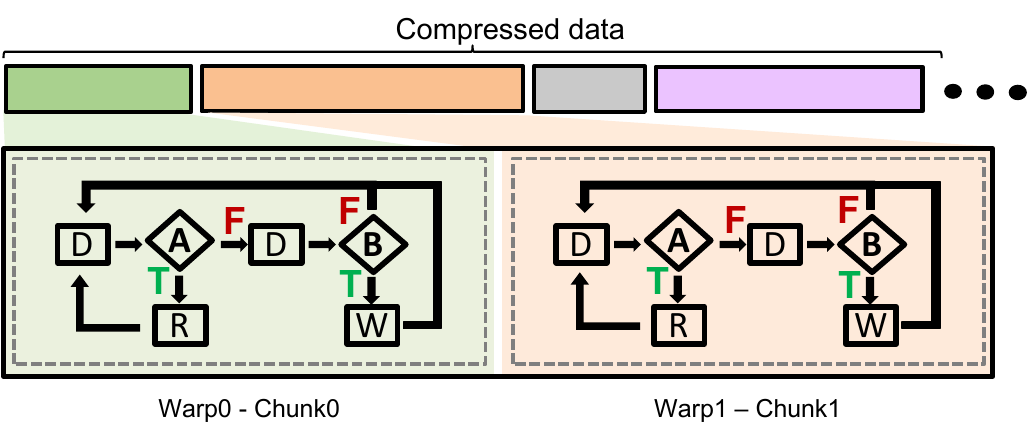}
        \vspace{-1ex}
        \caption{CODAG} 
        \label{fig:codagdesign}
     \end{subfigure}

     \includegraphics[width=1\textwidth]{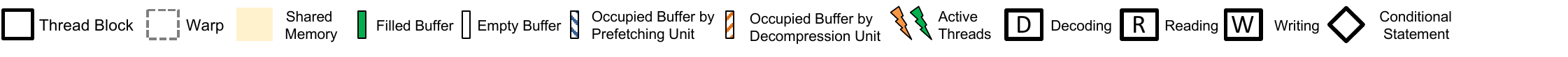}
           \vspace{-3ex}

    \setlength{\belowcaptionskip}{-10pt}
    \caption{Comparison of \pname{} with the state-of-the art decompressor NVIDIA RAPIDS implementation. In \pname{} design, the conditional statement \textbf{A} checks whether there is enough space in the input buffer to store full cacheline data, and \textbf{B} checks whether threads need to write uncompressed data based on the decoded information.
    }
      \vspace{-1ex}

}
\end{figure*}

\subsection{Parallelizing Decompression}
Designing high-throughput parallelizable decompression schemes is challenging due to the nature of compressed data.
First, compressors output a sequence of variable-length blocks of compressed data, often without any alignment guarantees (See $\S$~\ref{sec:comp_techniques}), resulting in size irregularities. 
Second, the encoding of a bit or byte in a compressed stream depends on the actions taken for the bits or bytes before it in the stream, creating data-dependencies in the compressed stream.
Thus, decompression requires starting at the first byte of the compressed stream and decoding the stream sequentially,
limiting parallelism during decompression.

To expose data parallelism during decompression, modern compression data formats 
divide the uncompressed input into fixed-size chunks, compress each chunk
independently, and keep the bytes of the compressed chunk sequential in memory~\cite{orc, parquet, rapids}. 
During decompression, each compressed chunk is assigned to a processing unit so that it can be decompressed independently of other chunks. 
The data format further specifies metadata for offsets in the uncompressed and/or compressed streams.

\subsection{State of the Art in Exploiting GPUs for Decompression}
\label{sec:backbaseline}
As GPUs become increasingly adopted by big-data applications due to their high compute throughput and memory bandwidth, there has been growing interest in high-performance decompression on GPUs. 
Generally, high-performance decompressors on GPUs assume modern data formats that support chunking.
Let's take a look at how the decompression kernel is mapped to GPU resources in the state-of-the-art NVIDIA RAPIDS implementation~\cite{rapids, nvcomp, nvcompblog}.

As shown in Figure~\ref{fig:baselinedesign}, after finding the offset for each compressed chunk, RAPIDS assigns each compressed chunk  
to a thread block, referred to as a decompression unit. 
Each decompression unit consists of a prefetching warp, batch buffers in the fast shared memory, and decoding/writing warps. 
RAPIDS opts to use a specialized prefetching warp that fetches bytes of the compressed stream from the slow global memory in a coalesced manner to the batch buffers. 
Having a specialized prefetching warp in the same thread block as the decoding/writing warps has the benefit of allowing the batch buffers to reside in the fast shared memory and the prefetch operations to overlap with the decoding operations for the same chunk. Both features help to reduce the latency of decompressing each chunk.

In the decoding warp, a leader thread is responsible for decoding the bytes from the compressed stream, as this is a sequential process. 
During decoding, the leader thread reads the data from the batch buffers. 
When the information has been decoded, the leader thread synchronizes with the other threads in the thread block and directs them to collectively perform a writing operation to the output decompressed data stream in a coalesced manner.


 \section{Characterization Study of State-of-the-art Decompressor}
\label{sec:charstudy}

Understanding and optimizing decompression on GPUs is essential as the decompression can consume approximately 91\% of total GPU time on a modern data analytics pipeline.
However, none of the prior works have performed a detailed analysis of existing state-of-the-art decompression schemes on GPUs.
To this end, we conduct an in-depth characterization study of the state-of-the-art decompressor in NVIDIA RAPIDS~\cite{rapids}, and 
identify its key performance bottlenecks. 


\textbf{Setup:} 
We use a variety of datasets across different domains as shown in Table~\ref{tab:workload} and evaluate RAPIDS's decompressor with RLE v1 and Deflate schemes. 
Due to space limitations, we only present the results from two datasets (\texttt{MC0} and \texttt{TPC}) but similar trends are observed for the rest.
To characterize decompressor performance, we use the system configuration described in Table~\ref{tab:config} and use the NVIDIA Nsight profiling tools to capture important metrics.

\begin{figure}[t]
  \includegraphics[width=\columnwidth]{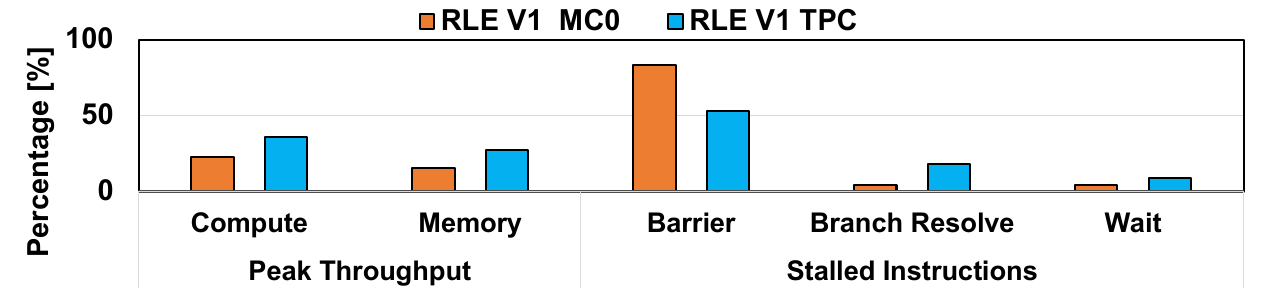}
  \vspace{-2ex}
  \setlength{\belowcaptionskip}{-15pt}
\caption{The peak throughput percentages and the stalled instruction distribution of RAPIDS decompressor for RLE v1.}

\label{fig:Baseline_Percentage}
\end{figure}

\textbf{Key insights:}
From the left side of Figure~\ref{fig:Baseline_Percentage}, we know that for the RLE v1 encoding scheme, both compute and memory resources are far from being fully utilized. As we discussed in $\S$\ref{sec:backbaseline}, these results point to latency exposed due to insufficient parallelism during the decoding process as all threads in a thread block wait on a barrier for the leader thread to complete its decoding operation. 
%
To understand the impact of the barrier, we measured the distribution of conditions causing an SM scheduler stall cycle, 
which is shown on the right side of Figure~\ref{fig:Baseline_Percentage}.
This metric provides the percentage of threads waiting for each type of stalling condition during the cycles where an SM scheduler cannot find a warp ready for execution. 
 
From Figure~\ref{fig:Baseline_Percentage}, during the SM scheduler stall cycles, most
(up to 80\%) of the threads are not ready because they are waiting on a barrier for the leader thread to complete its current decoding operation.
Apart from barriers, 
up to 20\% of the instructions are waiting for the resolution of branch conditions (Branch Resolve) when there is a lack of ready threads for the SM scheduler to dispatch for execution. 
The Wait metric indicates that a warp is stalled waiting on a fixed latency (arithmetic/logic unit) execution dependency. 
Moreover, some threads do not receive any work during the writing stage as run lengths can be smaller than the size of the thread block, further degrading the achievable performance.

\begin{figure}[t]
   \vspace{-2ex}
  \includegraphics[width=\columnwidth]{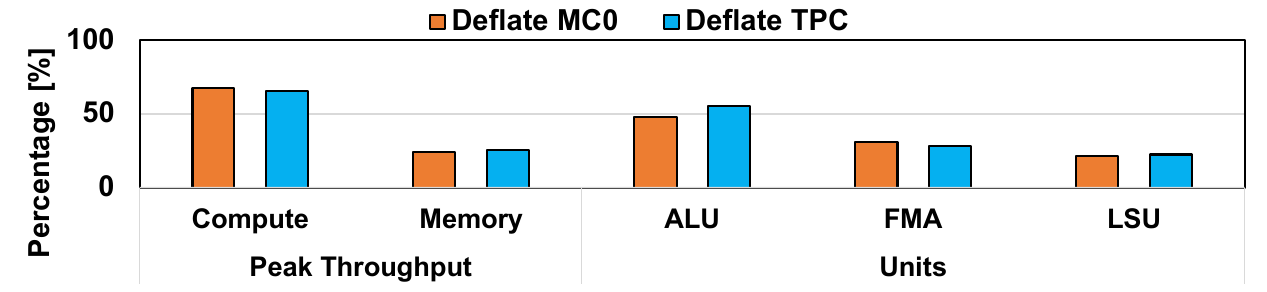}
\vspace{-3ex}
\caption{ The compute/memory peak throughput percentages and compute unit utilization of RAPIDS decompressor for Deflate.}
\vspace{-2ex}

\label{fig:Rapids_Deflate}
\end{figure}

For Deflate, unlike the traditional view that it is memory bandwidth bound~\cite{rapids, gompress}, our profiling results in Figure~\ref{fig:Rapids_Deflate} show that its memory bandwidth utilization is less than 25\%, 
which indicates that it is not limited by memory bandwidth but is compute limited. 
In contrast to RLE v1, Deflate has a much higher utilization of compute throughput (up to 65\%) than RLE v1 because of its high computational complexity in dictionary look-up and Huffman tree traversal. 
We need more detailed metrics and analysis in order to discern whether the performance is limited by a particular type of compute pipeline, memory latency, or compute latency.

We use the profiler to analyze the utilization of the different GPU compute pipelines for Deflate decompression.
The right side of Figure~\ref{fig:Rapids_Deflate} shows that the arithmetic logic units (ALU) are up to 55\% utilized while the fused multiply/add and load-store units are up to 35\% and 20\% utilized. 
This modestly high degree of ALU and FMA utilization mainly comes from the fact that during decoding, the leader thread executes a large number of arithmetic instructions for every byte to identify the next action for generating the output for the byte. 

Nevertheless, 
all these results point to memory latency and/or compute latency as the main limiter of Deflate performance. 
Due to the high computational complexity of Deflate decoding process, it takes a long time for the leader thread to complete the decoding of each byte while most threads in the thread block are waiting for the decoding thread. 
During this time, there are not enough decoding threads for the SM scheduler to tolerate the compute pipeline latency. As a result, the execution speed of Deflate is limited by compute pipeline latency in addition to barrier stalling.

\textbf{Summary and take-aways:}
From the profiling results, we see that the decompression speed of RLE v1 is limited by barrier latency and that Deflate is limited by barrier stall latency as well as the fixed latency of compute pipelines. Neither the large number of writing threads nor the specialized prefetching warps help address this limitation. Any further improvement in decompression speed for both RLE v1 and Deflate would require more simultaneously running decoding threads (more chunks being decoded in parallel), which is an important design goal of the \pname{} architecture.

\section{~\pname{} Design}
\label{sec:design}


The goal of \pname{} is to provide a flexible framework for high-throughput decompression on GPUs without requiring developers to implement sophisticated, GPU-specific optimizations.
\pname{}'s architecture, resource provisioning strategy, and kernel APIs enable developers to easily incorporate their encoding techniques of choice into \pname{} and rely on the rest of the framework to achieve high decompression speed. On the one hand, as discussed in Section~\ref{sec:charstudy}, the key to improved decompression speed is 
to increase the number of simultaneously running decompression threads so that the SM scheduler can better hide the synchronization/memory/compute-pipeline overheads.

On the other hand, it is also critical to maintain the coalesced access patterns in reading uncompressed data, reading dictionary value sequences, and writing decompressed data so that the memory access cost does not become excessive and creates new bottlenecks. 
Thus, the \pname{} architecture must address three key challenges in providing an efficient and effective solution that can be used with a wide variety of encoding techniques:

\begin{itemize}
    \item 
    \pname{} must provision GPU execution resources so that many threads can simultaneously decode more chunks in each SM to effectively tolerate latencies and efficiently utilize hardware resources.
    \item As the naïve on-demand reading and writing operations can make poor use of memory bandwidth and exhibit excessive latency, \pname{} must  optimally coalesce these accesses to minimize memory access overhead in spite of the irregular consumption of compressed data and sporadic generation of decompressed data by the decoder. 
    \item As the sophisticated optimizations for efficiently using memory bandwidth and tolerating latencies require significant programming skills and efforts, 
    the optimizations in the \pname{} framework must be generalized into reusable primitives for a wide range of encoding techniques.
\end{itemize}

\subsection{\pname{} System Overview and Abstraction}

Figure~\ref{fig:codagdesign} shows an overview of the \pname{} framework. 
\pname{} uses warps as decompression units instead of the larger thread blocks used by the state-of-the-art decompressor, enabling \pname{} to simultaneously decompress many compressed chunks.
This enables the hardware scheduler to overlap instruction execution from several independent warps and efficiently hide compute pipeline and memory latencies. 
In \pname{}, all threads in each warp perform the decoding operations as well as amalgamate to read uncompressed data and write decompressed data. 
Thus, \pname{} avoids the high barrier-synchronization overhead that the state-of-the-art decompressor incurs.

In \pname{}, each warp executes a decompression GPU kernel that implements a control flow of three main operations: sequential decoding, coalesced reading, and coalesced writing, as shown in Figure~\ref{fig:codagdesign}. 
All threads in a warp follow the same control flow to avoid using any broadcast synchronization operations. 
The control flow consists of a main loop for the decoding process and two conditional statements for on-demand reading and writing operations.  
During decoding, all threads sequentially execute the same decoding process in-compliance with the serial nature of the compressed data stream. 
In our current framework implementation, the sequential decoding code for different combinations of pertinent encoding techniques can be easily incorporated into the kernel as a CUDA device function.

For coalesced on-demand reading operations, all threads in the warp collaborate to fetch the next cache-line section (e.g., 128 bytes in the A100 GPUs) of the compressed input data stream in a coalesced manner into an input buffer. 
Having a buffer for the fetched bytes minimizes the number of fetches in spite of the potentially irregular patterns of compressed data consumption during decoding, which helps to reduce the cost of accessing the compressed data. 
In comparison with the RAPIDS decompressor, \pname{} does not have a specialized prefetch warp in each thread block to asynchronously prefetch input data while the thread block's leader thread is decoding. 
However, with many independent warps being simultaneously executed, \pname{} can still tolerate the memory latency.

When a writing operation is required during decoding, all threads in the warp collaborate to execute the coalesced on-demand writing of a cache-line of the uncompressed output data to the output buffer. 
Compared to the RAPIDS decompressor, this design at first glance may not seem to offer as much memory parallelism when writing into each output chunk, as RAPIDS uses thread blocks with up to 1024 threads available for writing into each output chunk.

However, the benefit with \pname{} is that warp-level barriers are significantly less expensive than their block-level counterparts. 
Furthermore, with the same thread block size, \pname{} supports decompressing and writing up to $32 \times$ more independent chunks in parallel. 
The increased data parallelism can easily compensate for the loss of memory-level parallelism within a chunk.  

As the reading pattern is the same across all encoding techniques, the on-demand reading operation code is portable to all (combinations of) encoding techniques. 
However, the on-demand writing operations can vary  depending on the encoding techniques and need to be specialized by decompressor developers. 
To shield the decompressor developers from the complexity of GPU algorithmic optimizations that are critical to high-performance writing operations, \pname{} provides several major types of optimized writing primitives to support a wide range of popular encoding techniques.

\begin{figure}[t]
{
\vspace{-4ex}
 \includegraphics[width=\columnwidth]{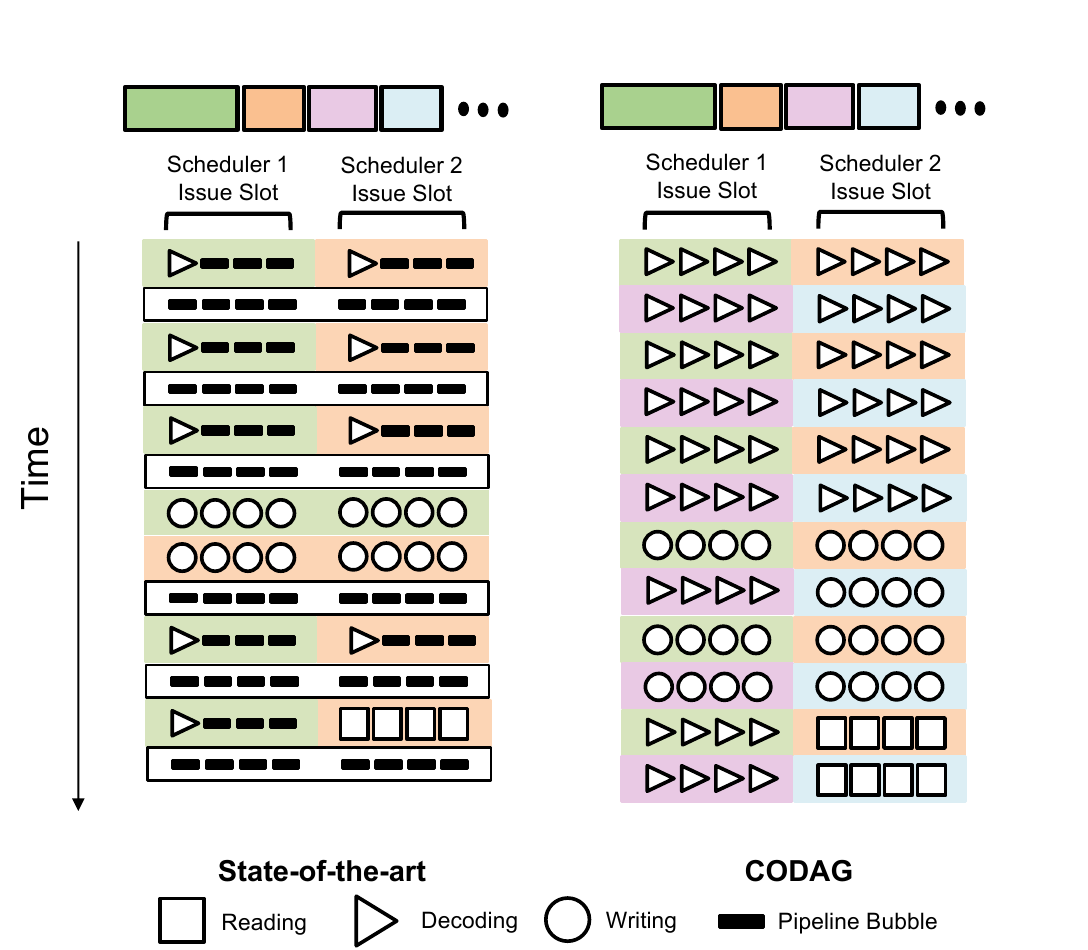}
 \setlength{\belowcaptionskip}{-10pt}
\vspace{-3ex}
  \caption{
Illustration of the issued instructions between the baseline and \pname{} in a steady state. 
 \label{fig:pipeline}
 }
\vspace{-0.5ex} 
}
\end{figure}

\subsection{\pname{} Meta-data and Performance APIs}
\lstset{language=c++,basicstyle=\footnotesize}       

\pname{} provides an infrastructure for realizing the warp-level decompression as well as the optimized reading and writing operations to lower the programming effort when using \pname{} for different encoding techniques. 
To isolate the complexity of GPU optimizations from decompressor developers, \pname{} provides \texttt{input\_stream} and \texttt{output\_stream} abstractions that transparently manage the synchronization, coalescing and metadata for the on-demand reading and writing operations. 
Developers can simply call the member functions, shown in Table~\ref{tab:Read_API} and Table~\ref{tab:Write_API}, as needed to read and write data efficiently.

To support the highly efficient on-demand reading and writing operations,  \texttt{input\_stream} and \texttt{output\_stream} internally manage meta-data such as offsets for the reading and writing pointers, information of input buffer, or fetched bites in the top entry of the input buffer. With these provided reading and writing primitives, \pname{} only requires small changes to the sequential decoding code as the memory access optimizations are hidden via the internal functions. 
In the next few sections, we will discuss the \pname{} decompression scheme and its optimizations more in detail.

{\renewcommand{\arraystretch}{1.2}
\begin{table}[t!]
\centering
\scriptsize
    \caption{\small \pname{} APIs for \texttt{input\_stream}.
}
\begin{tabular} {|p{1in}|p{2in}|}
\hline
\textbf{Member Functions} & \textbf{Description}\\
\hline
 \texttt{fetch\_bits(int n)} & {Fetch the next n bits in the compressed stream.}\\
\hline
 \texttt{peek\_bits(int n)} & {Peek at the next n bits in the compressed stream.}\\
\hline

\end{tabular}
\vspace{-2ex}
\label{tab:Read_API}
\end{table}
}

{\renewcommand{\arraystretch}{1.2}
\begin{table}[t!]
\centering
\scriptsize
    \caption{\small \pname{} APIs for \texttt{output\_stream}.
}
\begin{tabular} {|p{1.4in}|p{1.6in}|}
\hline
\textbf{Member Functions} & \textbf{Description}\\
\hline
 \texttt{write\_byte(int8\_t b)} & {Write a literal (byte) to the output buffer}\\ 
\hline
 \texttt{write\_run(T init, size\_t len, T delta)} & {Write a run based on an initial value of type \texttt{T}, a length, and a delta.} \\ 
\hline
\texttt{memcpy(int off, int len)} & {Memcpy operation for dictionary-based encoding with the offset and length.}\\
\hline

\end{tabular}
\vspace{-4ex}
\label{tab:Write_API}
\end{table}

}

\subsection{Warp Level Decompression}
\label{sec:design_warp}

The large granularity of the decompression units in the state-of-the-art limits the scheduler's ability to hide latency, because the vast majority of threads end up spending most of their execution time waiting for a few leader threads to finish sequential decoding. 
The left side of Figure~\ref{fig:pipeline} is a cartoon illustration of how the instructions are issued during the decompression for the state-of-the-art decompressor.  
We assume an SM with two schedulers and an occupancy capacity of 4 warps for the example. 
In this tiny example, assume that the state-of-the-art has a decompression unit of a thread block that consists of two warps. 
Thus, only two independent compressed chunks (green and orange) can be assigned to a single SM in the state-of-the-art where one of the two warps is specialized to prefetch data from global memory into the buffers, and both warps work together to perform writes to global memory after the leader has decoded a symbol.

We assume that during the first part of the timeframe, each scheduler is issuing instructions of the warp containing the leader thread of one of the decompressor units and, for simplicity, that the batch buffer is full so there is no prefetch activity. Since there is effectively only one thread active in each scheduler, the latency of the function units is fully exposed, shown as pipeline bubbles between the decoding operations. 
After the decoding operations are complete, all threads synchronize, incurring the synchronization pipeline bubbles before the write operations. 
The SM execution resources are underutilized due to the pipeline bubbles.

The right side of Figure~\ref{fig:pipeline} shows the execution timing of \pname{}. \pname{} uses warps as a decompression unit to allow more chunks (green, orange, purple, blue) to be decompressed simultaneously. 
This gives the hardware scheduler more independent warps to work with. 
Assume that the green and purple warps are assigned to scheduler 1 whereas the orange and blue are assigned to scheduler 2.  During the first part of the time frame, the decoding operations of the green and blue warps can be issued alternatively to tolerate each other's latency. Thus, the SM execution resources are much better utilized.  In reality, the reading, decoding, and writing operations of all the warps will be mixed together to more fully utilize the memory bandwidth than what is shown in 
Figure~\ref{fig:pipeline}. That is, since there are no inter-warp dependencies in \pname{}, the SM's scheduler has a wider pool of useful instructions to schedule from and thus can better reduce pipeline bubbles and hide latencies. Furthermore, there is no more synchronization overhead before the writing operations.

\subsection{\pname{} Decoding Scheme}
\label{sec:decoding_scheme}



As the prefetch warp is removed in \pname{}'s warp-level decompression scheme, all threads in the warp collectively execute on-demand reading operations during decoding whenever there are not enough fetched bytes. To do this, 
the decoding state needs to be saved before the reading operation and restored after the reading operation to enable the leader thread to continue decoding after the reading operations. 

To avoid the above-mentioned overhead during the transitions between the decoding operations and the reading/writing operations, \pname{} uses an all-thread decoding technique by activating all threads in the warp to execute the same serial decoding operations on the same chunk. 
In this decoding scheme, all threads have a local copy of decoded information as they execute the same decoding process thus broadcasts are no longer needed. 
As for the reading operations, all threads in the warp only need to perform a low-cost warp-level synchronization to ensure a consistent view of the input buffer before 
reading as newer GPU architectures~\cite{voltawhite,amperewhite} do not guarantee lock-step execution of the threads in a warp.
After the reading operation, all threads in the warp can return naturally to continue the decoding operation, 
eliminating the need for explicitly saving and restoring the decoding state.

Since the decoding operation is redundantly performed by all threads in a warp, one might take a pessimistic view of the all-thread decoding technique. Thus, we use a micro-benchmark to compare the ALU compute throughput of the single-thread decoding and the all-thread decoding. For this micro-benchmark, the number of arithmetic operations executed per global memory access is varied from 1 to 100,000 to capture the compute throughput at different compute intensity levels. 
The result shows that the difference in compute throughput of the two decoding techniques never exceeds 0.1\%. Thus, the additional arithmetic operations executed due to the all-thread decoding technique do not degrade the achieved compute throughput.

\begin{algorithm}
\scriptsize
\caption{\small On-demand Reading With Shared Input Buffer}\label{alg:reading}
\begin{algorithmic}[1]
\IF{len - counter $<$  128}
    \STATE  barrier()
    \STATE  read\_idx = \text{offset} + lane * 4
    \STATE  enq\_idx = (head + counter + lane) \% len
    \STATE  \text{Buff}[enq\_idx]  = \text{Input}[read\_idx]
    \STATE  counter += 128
    \STATE  barrier()
\ENDIF

\STATE req\_val = \text{Buff}[$head$]
\STATE head = (head + 1)  \% len
\STATE counter-{}-
\RETURN req\_val

\end{algorithmic}
\end{algorithm}

\subsection{Coalesced On-Demand Reading Operation}
To maximize the benefits of the warp-level decompression scheme,~\pname{} also provides the APIs for highly efficient on-demand reading operations since
it is also critical to maintain the coalesced access patterns in reading compressed data to optimize GPU memory bandwidth utilization. 
Thus, \pname{} implements coalesced on-demand reading operations behind the \texttt{input\_stream} abstraction.

\pname{}’s on-demand reading operation with its default configuration is illustrated with the pseudocode in Algorithm ~\ref{alg:reading}. By default, \pname{} deploys an input buffer in shared memory, to temporarily store the fetched bytes, minimizing the number of global memory accesses. The size of the input buffer is at least twice the cacheline size so that entire cachelines can be easily enqueued in the input buffer. An input buffer also has a head pointer for threads to determine where to read from in the input buffer, and a length to keep track of the number of bytes left in the buffer.

For a given on-demand reading operation, all threads in a warp first collectively refill the shared input buffer with bytes fetched from a cacheline. Each thread in the warp fetches 4-bytes after a warp-synchronization barrier to ensure coalesced memory accesses. Once the input buffer is refilled, threads update the offset into the input in the \texttt{input\_stream} instance. Then, the threads read the requested bits from the input buffer using the head pointer and update the internal state of its \texttt{input\_stream} for the next on-demand reading operation. 

\textbf{Using Registers:} The shared memory usage for the input buffer might limit the applicability of \pname{} in applications with high shared memory utilization. To address this, \pname{} provides a configuration where the input buffer is implemented with the threads' local registers instead of the shared memory. In this configuration, each thread uses two 32-bits registers to implement the input buffer.

Similar to the default configuration, the input buffer is first refilled when there is enough space to store a complete cacheline. 
This is done by leveraging double buffering where two sets of 32 local registers across the warp operate as a double buffer. 
Since the data is stored in a local register, \pname{} exploits warp-wise broadcast when fetching requested bytes from the input buffer. 
For this operation, the head pointer is first used to identify which thread’s local register is holding the requested bytes so that the requested bytes can be broadcast by the appropriate thread to all other threads in a warp.

\subsection{Coalesced On-demand Writing Operation}

\begin{algorithm}
\scriptsize
\caption{\small Memcpy(offset, len)}\label{alg:memcpy}
\begin{algorithmic}[1]

\IF{{!is\_4B\_aligned(write\_ptr)}}
    \STATE {pad\_bytes = (4 - (write\_ptr \% 4)) \% 4}
    \STATE {byte\_memcpy(offset, pad\_bytes)}
\ENDIF
\STATE {barrier()} \
\STATE {num\_itr = ceil((len - pad\_bytes) / 128)}\
\FOR{(int i = 0; i < num\_itr; i++)}
    \STATE {read\_idx = ceil((read\_ptr + lane*4) / 4)}
    \STATE {r1 = out[read\_idx-1]}
    \STATE {r2 = out[read\_idx]}
    \STATE {rdata = funnel\_shift(r1, r2, read\_ptr \% 4)}\
    \STATE {{out}[(write\_ptr + 4*tid) / 4]  = rdata}\
    \STATE {read\_ptr, write\_ptr += 32}
    \STATE barrier() 
\ENDFOR

\end{algorithmic}
\end{algorithm}

Similar to the on-demand reading operations, \pname{} supports highly efficient coalesced on-demand writing operations to reduce the cost of memory accesses. Ideally, all threads in a warp should collectively execute an on-demand writing operation of a full cacheline and then synchronize. 
However, as the required writing operations can vary based on the encoding techniques, \pname{} provides optimized writing primitives to support the most common encoding techniques: 1) writing a literal value, 2) writing a run with delta (delta is 0 for a simple run), and 3) memory copy (memcpy).

\textbf{Writing a literal value:} As this operation only requires a few bytes to write to the output buffer, only one thread executes this operation.

\textbf{Writing a run}: This operation is commonly used for RLE-based encoding techniques. For this operation, each thread independently computes its decompressed value based on the initial value, delta, and thread index and then writes it to the output buffer at the appropriate offset to maximize memory utilization.  

\textbf{Memcpy}: Common dictionary-based encoding techniques compress data with a length specifying the number of bytes to copy from the dictionary to the output buffer and an offset into the output buffer from the last byte written before the current memcpy starts.
Unlike a regular memcpy, using a warp to perform a memcpy operation in a dictionary-based compression algorithm has a couple of challenges. 

First, as the dictionary is implicit in the previously written output, the reading and the writing pointer both point to the same output buffer. Second, common algorithms only provide byte alignment for both the input and the output.
These characteristics make it difficult to exploit the warp's parallelism in memcpy, especially when we would like each thread to copy more than 1-byte (e.g. 4-bytes) at a time for better coalescing, reduced executed memory instructions, and improved GPU cacheline and memory bandwidth utilization.

To address these challenges, we design \pname{}’s fast memcpy operation, shown in Algorithm~\ref{alg:memcpy}. 
For a given memcpy operation, the threads first fetch the writing pointer from \texttt{output\_stream} and check whether it is 4-byte aligned (line 1 in Algorithm~\ref{alg:memcpy}). 
If it is not aligned, the threads collectively copy up to 3 bytes to guarantee 4-byte alignment (line 3). Once the writing pointer is aligned, the warp synchronizes to ensure that the copied bytes are visible to the whole warp. 
Then, the warp executes copy operation in iterations to cover the entire memcpy length, with each thread writing one 4-byte value to the output each iteration. Each thread independently calculates its reading pointer for a given iteration based on its warp lane index (line 8). 
Since the reading pointer is not guaranteed to be 4-byte aligned, each thread reads two consecutive 4-byte elements (line 9-10), then bit-shifts and concatenates them to generate the correct 4-byte output element (line 11) to write to its assigned offset (line 12). 

Note, in each iteration, each thread in the warp independently reads two 4-byte elements written in previous iterations and writes one complete 4-byte element. 
We found that using 4-byte granularity for the memcpy loop body provided the best trade-off between overhead to align the writing pointer, the amount of parallelism exposed, and cacheline utilization. All threads in a warp synchronize at the end of each loop iteration, preventing race conditions between iterations.

Moreover, with dictionary compression, it is possible that the length of a memcpy is larger than the offset. 
In this special case, the byte sequence from the offset till the end of the output (before the memcpy executes) is to be appended to the output, possibly multiple times.
Just like in Algorithm ~\ref{alg:memcpy}, we first align the writing pointer to 4-bytes and then assign each thread one 4-byte output element to generate per memcpy loop iteration. 
However, in this special case, each 4-byte output element is created by combining two consecutive 4-byte elements in a fixed circular window of 4-byte elements, starting from the 4-byte element pointed to by the offset until the last 4-byte element of the output buffer. 
Thus, each thread can independently use modulo-arithmetic to find the two consecutive 4-byte elements in the specified window, and bit-shift and concatenate them to generate its output element each iteration. 
We don't show the algorithm listing for this case due to space constraints.

\section{Evaluation}
\label{sec:eval}
Our evaluation demonstrates that: 
(1) ~\pname{} framework improves the performance of decompression by efficiently utilizing the hardware resources provided by the GPU. ~\pname{} provides 13.46$\times$, 5.69$\times$, 1.18$\times$ decompression throughput speed up (geomean) on ORC RLE v1, ORC RLE v2, and Deflate compared to the state-of-the art implementation in NVIDIA RAPIDS~\cite{rapids}. (2) A decompressor implemented in the ~\pname{} framework can rival the state-of-the-art GPU implementation without application-specific optimizations. 

\subsection{Experiment Setup}
We evaluate \pname{} using three different encoding techniques, namely RLE v1 and RLE v2 from the Apache ORC file format, and Deflate. 
These encoding schemes are chosen due to wide applicability but other encodings can also be enabled.
We compare \pname{} and the state-of-the-art NVIDIA RAPIDS baseline on an environment shown in Table~\ref{tab:config}.
We lock the GPU's clock frequency to peak before running the experiments.

\subsection{Evaluation Datasets}
For each algorithm, we evaluate \pname{} and the baseline on seven real-world sample datasets curated from four industrial application domains as shown in Table~\ref{tab:workload}. 
Each dataset has unique properties enabling us to evaluate the proposed scheme on not only different types of encoding techniques but also different data types and patterns. 
For example, \texttt{MC0} and \texttt{MC3} have long run-lengths while \texttt{TPC} and \texttt{TPT} contains many repeated patterns. 
\texttt{CD2}, and \texttt{TC2} follow power law distributions for the data. 
\texttt{HRG} is human genome dataset and has repeated text pattern containing \texttt{A},\texttt{T},\texttt{C},\texttt{G} and \texttt{N} characters. 

For the baseline, we use the official data management tools provided by ORC~\cite{orc} to compress data with RLE v1 and RLE v2 algorithms. To compress the data with deflate, we use the zlib library~\cite{zlib} with a compression level of 9, the highest compression level. The chunk size for the original data is fixed to be 128KB for both \pname{} and the baseline.

{\renewcommand{\arraystretch}{1.2}
\begin{table}[t]
\centering
\scriptsize
    \caption{\small {Configuration used to evaluate \pname{}.}}
    \vspace{-2ex}
\begin{tabular}{|p{0.55in}|p{2.25in}|}
    \hline
    {\textbf{Configuration}}& {\textbf{Specification}} \\
    \hline
    \hline
	{CPU}                  & {AMD EPYC 7702 64-Core Processor} \\
    \hline
	{Memory}               &  {1TB DDR4} \\
    \hline
	\multirow{1}{*}{GPU 1} &  NVIDIA Tesla V100 HBM2 32GB \\ 
    \hline
	\multirow{1}{*}{GPU 2} &  NVIDIA A100 HBM2 40GB \\ 
    \hline
\end{tabular}
\label{tab:config}
\end{table}
}

{\renewcommand{\arraystretch}{1.2}
\begin{table}[t!]
\centering
\scriptsize
    \caption{\small Real-world dataset used for evaluating \pname{}.
}

\begin{tabular}{|p{1.3in}|p{0.6in}|p{0.3in}|p{0.40in}|}
    \hline
    \textbf{Dataset} & \textbf{Category} & \textbf{DType}& \textbf{Size (GB)}  \\
    \hline
    \hline
	Mortgage Col 0 (MC0)~\cite{mortgage} & Analytics & uint\_64  & 4.86  \\
    \hline
	Mortgage Col 3 (MC3)~\cite{mortgage} & Analytics & fp32    & 2.43  \\
    \hline
	NYC Taxi Passenger Count (TPC)~\cite{nyctaxi} & Analytics & int\_8  & 3.07  \\
    \hline
	NYC Taxi Payment Type (TPT)~\cite{nyctaxi} & Analytics & char  & 7.41  \\
    \hline
    Criteo Dense 2 (CD2)~\cite{criteo} & Recommenders & uint\_32  & 0.73  \\
    \hline
    Twitter COO Col 1 (TC2)~\cite{twitter} & Graph & uint\_64  & 5.47  \\
    \hline
    Human Reference Genome (HRG)~\cite{genomics} & Genomics & char  & 3.1  \\
    \hline

\end{tabular}
\label{tab:workload}
\vspace{-2ex}
\end{table}
}

{
\begin{table}[t!]
\scriptsize
\begin{center}
 \caption{\small Compression Ratios and average compressed symbol length when the datasets are compressed by RLE v1, RLE v2, and Deflate. The compressed symbol length for RLE v2 is excluded due to various compressed types for RLE v2. }
\begin{tabular}{|p{0.3in}|p{0.4in}|p{0.4in}|p{0.4in}|p{0.4in}|p{0.4in}| }
   \hline
   \multirow{2}{*}{\textbf{Dataset} } & \multicolumn{3}{c|}{\textbf{Compression Ratio}} & \multicolumn{2}{c|}{\textbf{Avg Comp Sym Len}} \\
   \cline{2-6}
   & \textbf{RLE v1} & \textbf{RLE v2} & \textbf{Deflate} & \textbf{RLE v1} & \textbf{Deflate} \\
    \hline
    \hline
    MC0 & 0.023 & 0.022 & 0.017 & 29.7 & 81.3 \\
    \hline
    MC3 & 0.038 & 0.039 & 0.015 & 40.5 & 156.6 \\
    \hline
    TPC & 0.867 & 0.637 & 0.119 & 1.00 & 10.1 \\
    \hline
    TPT & 1.41 & 0.99 & 0.042 & 1.00 & 32.6 \\
    \hline
    CD2 & 0.286 & 0.308 & 0.625 & 20.9 & 5.89 \\
    \hline
    TC2 &  0.087 & 0.075 & 0.0172 & 34.3 & 67.3 \\
    \hline
    HRG & 0.975 & 0.972 & 0.305 & 1.00 & 7.76 \\
    \hline
\end{tabular}
\end{center}

 \vspace{-2ex}
\label{tab:dataset_info}
\end{table}
}

\begin{figure}[t]
  \includegraphics[width=\columnwidth]{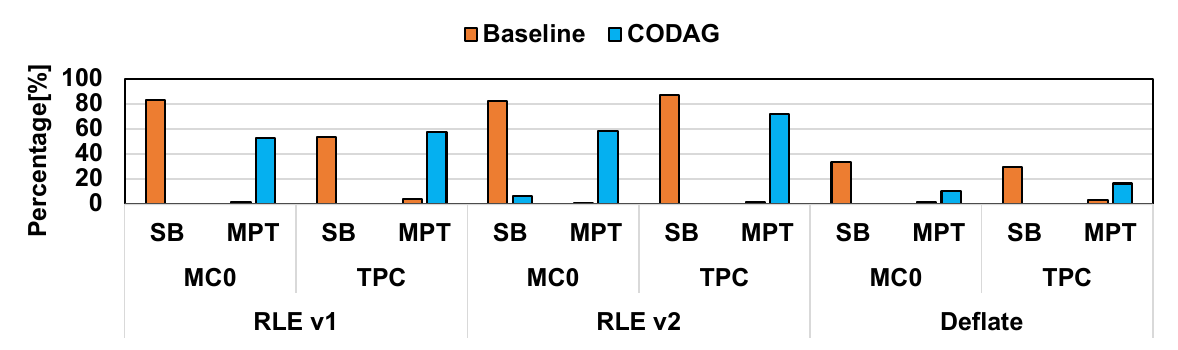}
  \caption{Stalled instruction distribution comparison between \pname{} and RAPIDS baseline.}
  \vspace{-2ex}
\label{fig:CODAG_Inst}
\end{figure}

\begin{figure}[h]
  \includegraphics[width=\columnwidth]{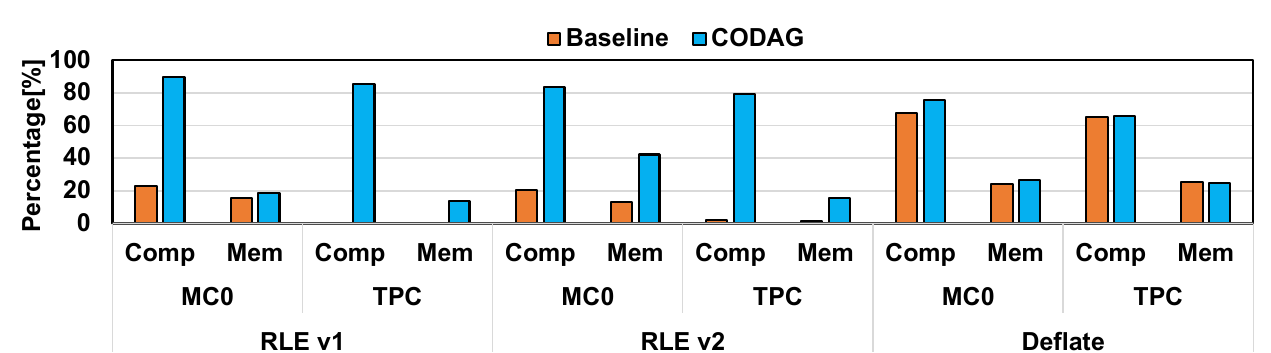}
  \caption{\small Comparing the compute/memory peak throughput of \pname{} and RAPIDS. Comp/Mem stand for compute/memory throughput percentage.}
\vspace{-4ex}
\label{fig:CODAG_Analysis}

\end{figure}

\subsection{Latency Analysis} 

To analyze how effectively \pname{} hides latency and utilizes resources, we measured the stalled instruction distribution and compute/memory throughput percentages of \pname{} and the baseline. 
For the stalled instruction distribution, we measured the number of instructions stalled due to synchronization (SB) and Math Pipeline Throttle (MPT). Measuring SB helps to understand thread occupancy during decompression, and measuring MPT helps us to understand the pressure on computing units as MPT means the instructions are stalled to wait on a specific oversubscribed math pipeline.

In this experiment, \texttt{MC0} and \texttt{TPC} datasets are used for the measurement as they have drastically different compression ratios. As shown in Figure~\ref{fig:CODAG_Inst}, we observe a dramatic difference in SB percentages between the baseline and~\pname{}. 
The lower SB percentage shows that there are more active threads during decompression.
Moreover, we observe substantially more instructions are stalled due to MPT in~\pname{}, showing~\pname{} utilizes more compute resources
and shifts decompression from latency-bound to compute-bound on GPUs.

Figure~\ref{fig:CODAG_Analysis} shows that \pname{} achieves a higher memory throughput percentage compared to the baseline. This is because \pname{} optimally utilizes memory bandwidth by leveraging coalesced on-demand reading/writing operations and enabling more memory instructions to be executed in parallel by higher ILP.

\begin{figure*}[t]
\begin{centering}
  \includegraphics[width=\textwidth]{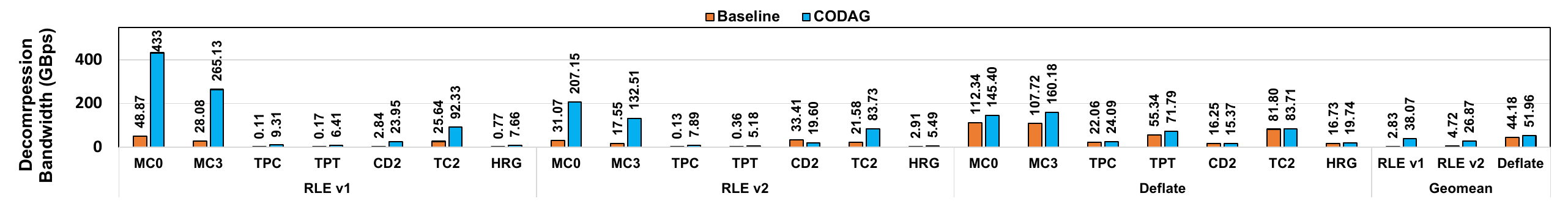}
\setlength{\belowcaptionskip}{-10pt}
\vspace{-4ex}
\caption{Comparing decompression throughput of \pname{} and RAPIDS baseline on A100 GPU.}
\label{fig:Decompression_Bandwidth}
\end{centering}
\end{figure*}

\begin{figure*}[t]
\begin{centering}
  \includegraphics[width=\textwidth]{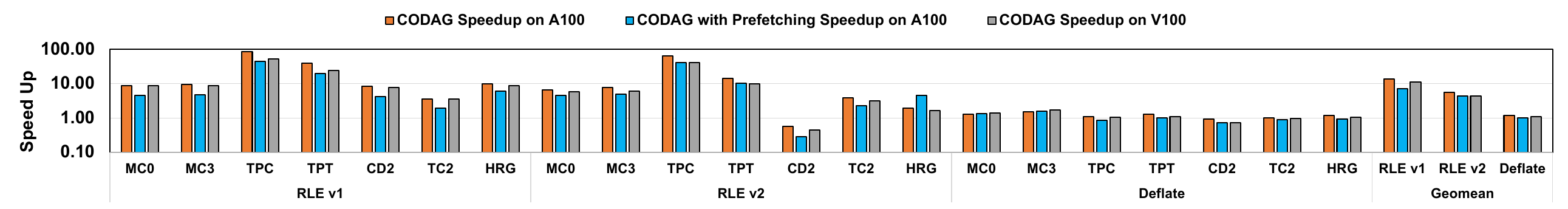}
  \setlength{\belowcaptionskip}{-10pt}

    \vspace{-3ex}
  \caption{\pname{} speedup without and with a prefetching warp when compared to RAPIDS in the latest two generations of NVIDIA GPUs. First two bars represent the speedup compared to RAPIDS on A100, and the baseline for the third bar is RAPIDS on V100.}
  \vspace{-1ex}

\label{fig:speedup}
\end{centering}
\end{figure*}

\subsection{Decompression Throughput} 
We now evaluate the decompression throughput (output bytes per second) of \pname{} and RAPIDS on three encoding techniques. Figure~\ref{fig:Decompression_Bandwidth} shows the decompression throughput of each encoding technique for both \pname{} and the baseline, and Figure~\ref{fig:speedup} shows the \pname{} speedup over the baseline decompressor. As shown in Figure ~\ref{fig:Decompression_Bandwidth}, \pname{} achieves 38.07 GBps, 26.87 GBps, and 51.96 GBps geo-mean decompression throughput while the state-of-the-art decompressor achieves 2.83 GBps, 4.72 GBps, and 44.18 GBps. Thus, \pname{} achieves 13.46$\times$, 5.69$\times$, and 1.18$\times$ geo-mean speed ups compared to the baseline as shown in Figure ~\ref{fig:speedup}.

\pname{} sees significantly higher decompression throughput in \texttt{MC0} and \texttt{MC3} datasets. This is because highly compressable datasets amplify the effective memory bandwidth, requiring fewer bytes to decompress the chunk and enabling more active threads during writing operations. In our datasets, the mortgage datasets (\texttt{MC0} and \texttt{MC3}) are highly compressable datasets, thus showing the highest decompression throughput. 

We note that \pname{} achieves substantially higher speedups over the baseline for RLE-based encoding techniques than Deflate. This is because with the state-of-the-art the average run length needs to be longer for RLE v1 for best thread occupancy; else, most threads become inactive during writing operations.

\subsection{Single-Thread Decoding vs All-Thread Decoding}

Next, we present the benefit of performing all-thread decoding over single-thread decoding for RLE v1 (RLE v2 has similar behavior) and Deflate decompression.
%
%
%
When we compare the decompression throughput of the two decoding techniques, the result shows the \pname{} with the all-thread decoding technique achieves 1.17$\times$ and 1.19$\times$  higher geo-mean decompression throughput than 
the single-thread decoding technique. 
As discussed in $\S$~\ref{sec:decoding_scheme}, this shows~\pname{} successfully reduces the number of 
broadcast operations while maintaining the achieved compute throughput of ALUs.

\subsection{Impact of Finer Decompression Unit}
This section presents an ablation study to show that (1)
reducing the number of threads responsible for decoding a chunk is key to \pname{}'s performance gains, and (2) \pname{}'s warp-level decompression, i.e., without a prefetching warp, further improves performance.

For this evaluation, we add a prefetch warp to \pname{}, such that two warps are scheduled per chunk - one warp to prefetch compressed input data and one warp to perform the decompression. This makes the \pname{} architecture more similar to the baseline except that the baseline allocates more threads to decompress each chunk - 1024 threads for RLE v1 and RLE v2 and 128 threads for Deflate.


Figure \ref{fig:speedup} shows \pname{} with a prefetching warp is able to achieve 7.10$\times$, 4.33$\times$, and 1.02$\times$ higher geo-mean decompression throughputs compared to the baseline for the three respective encoding techniques. 
Based on the result, finer decompression units enable the GPU hardware to tolerate compute operation and memory latency more effectively.  

However, the performance gains for each of the three encoding techniques is lower than the performance gains achieved by \pname{}'s single-warp decompression unit, i.e. without a prefetching warp. This is because the decompression process is not bounded by memory bandwidth, thus manually allocated hardware resources for prefetching to tolerate memory latency limits hardware resource utilization. Additionally, \pname{} minimizes the cost of memory access latency by coalesced on-demand memory operations. Thus, the benefits from reallocating the resources used for prefetching to decoding out-scales the benefits from prefetching units.

\subsection{Impact of Hardware Resources}

In the previous sections, we establish that \pname{} offers superior decompression bandwidth on A100 GPU compared to the baseline. In this section, we constrain the \pname{} and baseline design by downgrading the GPU to NVIDIA Volta V100. From this experiment, we analyze the impact of hardware resources on \pname{} and the scalability of \pname{} and the baseline design. 

As shown in Figure \ref{fig:speedup}, \pname{} achieves 11.19$\times$, 4.39$\times$, and 1.10$\times$ geo-mean speed up compared to the baseline on V100 while it achieves 13.46$\times$, 5.69$\times$, and 1.18$\times$ geo-mean speed up compared to the baseline on A100. The impact of compute under-utilization becomes more critical as newer GPU provides higher compute resources and memory bandwidth. Thus, we observe \pname{} scales better than the baseline with the available hardware resources as it more efficiently exploits parallelism provided by the GPUs. 

\section{Related Work}
\label{sec:related}

\textbf{Parallel compression/decompression with CPUs:} 
Multiple efforts to parallelize compression and decompression algorithms on CPUs have been proposed in the past~\cite{pigz, pbzip2, zstd}. 
pigz decompression is inherently hard to parallelize due to data dependency between the variable-length compressed blocks ~\cite{pigzreason, pigz}. Zstd~\cite{zstd} and pbzip2~\cite{pbzip2} implementations add decompressed size as part of their headers to exploit parallelism. 
However, such coarse-grain parallelism across the compressed blocks is insufficient to fully utilize GPUs~\cite{gompress}.



\textbf{GPU Decompression:} 
Several prior works have proposed implementing efficient compression or encoding algorithms for GPUs~\cite{rlecomp, nvcomp, bwtgpu,culss,gompress,fpcomp,pipelinecomp,texturecomp,all,tile_compression}. 
The prior works~\cite{gompress,all,tile_compression, rlecomp,fpcomp}, define their own file formats to achieve high decompression throughput for Deflate algorithm on GPU and propose to split the input data into smaller data blocks. However, compared to~\pname{}, the datasets should go through expensive pre-processing steps to recompress the entire datasets into their own data format. Other prior works~\cite{pipelinecomp, texturecomp,culss,bwtgpu} leverage algorithm-specific optimizations, limiting flexibility. 

\section{Conclusion}
\label{sec:conclusion}
In this work, we introduced \pname{}, a flexible framework for high throughput decompression on GPUs without requiring decompressor developers to implement sophisticated, GPU-specific optimizations.
\pname{} exploits warp-level decompression and highly efficient on-demand reading and writing operations to effectively hide latencies. 
Moreover, \pname{} completely removes the broadcast operations by dedicating all threads to participate in decoding. 
Our evaluations show that \pname{} can achieve 13.46$\times$, 5.69$\times$, and 1.18$\times$ speed up compared to the state-of-the-art decompressor for RLE v1, RLE v2, and Deflate encoding techniques.




\bibliographystyle{IEEEtran}
\bibliography{ref}

\end{document}